\documentclass[longauth,letter]{aa} 

\usepackage{graphicx}
\usepackage{txfonts}
\usepackage{color}

\usepackage{url}

\begin{document} 

   \title{The peak-flux of GRB 221009A measured with \textit{GRBAlpha}}
    
   \author{Jakub \v{R}\'{\i}pa\inst{1}
           \and
           Hiromitsu Takahashi\inst{2}
           \and
           Yasushi Fukazawa\inst{2}
           \and
           Norbert Werner\inst{1}
           \and
           Filip Münz\inst{1}
           \and
           Andr\'as P\'al\inst{3}
           \and
           Masanori Ohno\inst{2}
           \and
           Marianna Daf\v{c}\'{\i}kov\'a\inst{1}
           \and
           L\'aszl\'o M\'esz\'aros\inst{3}
           \and
           Bal\'azs Cs\'ak\inst{3}
           \and
           Nikola Hus\'arikov\'a\inst{1}
           \and
           Martin Kol\'a\v{r}\inst{1}
           \and
           G\'abor Galg\'oczi\inst{4,5}
           \and
           Jean-Paul Breuer\inst{1}
           \and
           Filip Hroch\inst{1}
           \and
           J\'an Hudec\inst{6}
           \and
           Jakub Kapu\v{s}\inst{6}
           \and
           Marcel Frajt\inst{6}
           \and
           Maksim Rezenov\inst{6}
           \and
           Robert Laszlo\inst{7}
           \and
           Martin Koleda\inst{7}
           \and
           Miroslav \v{S}melko\inst{8,9}
           \and
           Peter Han\'ak\inst{9}
           \and
           Pavol Lipovsk\'{y}\inst{9}
           \and
           Tom\'a\v{s} Urbanec\inst{10}
           \and
           Miroslav Kasal\inst{10}
           \and
           Ale\v{s} Povala\v{c}\inst{10}
           \and
           Yuusuke Uchida\inst{11}
           \and
           Helen Poon\inst{2}
           \and
           Hiroto Matake\inst{2}
           \and
           Kazuhiro Nakazawa\inst{12}
           \and
           Nagomi Uchida\inst{13}
           \and
           Tam\'as Boz\'oki\inst{14}
           \and
           Gergely D\'alya\inst{15}
           \and
           Teruaki Enoto\inst{16}         
           \and
           Zsolt Frei\inst{4}
           \and
           Gergely Friss\inst{4}
           \and
           Yuto Ichinohe\inst{17}
           \and
           Korn\'el Kap\'as\inst{18,19,5}
           \and
           L\'aszl\'o L. Kiss\inst{3}
           \and
           Tsunefumi Mizuno\inst{2}
           \and
           Hirokazu Odaka\inst{20}
           \and
           J\'anos Tak\'atsy\inst{4,5}
           \and
           Martin Topinka\inst{21}
           \and
           Kento Torigoe\inst{2}
          }

   \institute{Department of Theoretical Physics and Astrophysics, Faculty of Science, Masaryk University,  Kotl\'a\v{r}sk\'a 267/2, Brno 611 37, Czech Republic; \email{ripa.jakub@gmail.com} 
   \and
   Department of Physics, Graduate School of Advanced Science and Engineering, Hiroshima University, Higashi-Hiroshima, Japan 
   \and
   Konkoly Observatory, Research Centre for Astronomy and Earth Sciences, Budapest, Hungary 
   \and
   E\"otv\"os Lor\'and University, Budapest, Hungary 
   \and
   Wigner Research Centre for Physics, Budapest, Hungary 
   \and
   Spacemanic Ltd, Bratislava, Slovakia 
   \and
   Needronix Ltd, Bratislava, Slovakia 
   \and
   EDIS vvd., Ko\v{s}ice, Slovakia 
   \and
   Faculty of Aeronautics, Technical University of Ko\v{s}ice, Slovakia 
   \and
   Department of Radio Electronics, Faculty of Electrical Engineering and Communication, Brno University of Technology, Brno, Czech Republic 
   \and
   Tokyo University of Science, Noda, Chiba, Japan 
   \and
   Department of Physics, Nagoya University, Nagoya, Aichi, Japan 
   \and
   Institute of Space and Astronautical Science, Japan Aerospace Exploration Agency, Japan 
   \and
   Institute of Earth Physics and Space Science (EPSS), Sopron, Hungary 
   \and
   Department of Physics and Astronomy, Universiteit Gent, B-9000 Ghent, Belgium 
   \and
   School of Science, Kyoto University, Kyoto, Japan 
   \and
   Department of Physics, Rikkyo University, Tokyo, Japan 
   \and
   Department of Theoretical Physics, Institute of Physics, Budapest University of Technology and Economics, M\H{u}egyetem rkp. 3, H-1111 Budapest, Hungary 
   \and
   MTA-BME Quantum Dynamics and Correlations Research Group,  Budapest University of Technology and Economics, M\H{u}egyetem rkp. 3, H-1111 Budapest, Hungary 
   \and
   Department of Earth and Space Science, Osaka University, Toyonaka, Osaka, Japan 
   \and
   INAF - Istituto di Astrofisica Spaziale e Fisica Cosmica, Via A. Corti 12, I-20133 Milano, Italy 
   }

\offprints{J. \v{R}\'{\i}pa, \email{ripa.jakub@gmail.com}}
\date{Received date / Accepted date}


  \abstract
   {On 2022 October 9 the brightest gamma-ray burst (GRB) ever observed lit up the high-energy sky. It was detected by a multitude of instruments, attracting the close attention of the GRB community, and saturated many detectors.}
   {\textit{GRBAlpha}, a nano-satellite with a form factor of a 1U CubeSat, has detected this extraordinarily bright long-duration GRB 221009A without saturation, but affected by pile-up. We present light curves of the prompt emission in 13 energy bands, from 80\,keV to 950\,keV, and perform a spectral analysis to calculate the peak flux and peak isotropic-equivalent luminosity.}
   {Since the satellite's attitude information is not available for the time of this GRB, more than 200 incident directions were probed in order to find the median luminosity and its systematic uncertainty.}
   {We found that the peak flux in the $80-800$\,keV range (observer frame) was $F_{\rm{ph}}^{\rm{p}}=1300_{-200}^{+1200}$\,ph\,cm$^{-2}$s$^{-1}$ or $F_{\rm{erg}}^{\rm{p}}=5.7_{-0.7}^{+3.7}\times10^{-4}$\,erg\,cm$^{-2}$s$^{-1}$ and the fluence in the same energy range of the first GRB episode lasting 300\,s, which was observable by \textit{GRBAlpha}, was $S=2.2_{-0.3}^{+1.4}\times10^{-2}$\,erg\,cm$^{-2}$ or $S^{\rm{bol}}=4.9_{-0.5}^{+0.8}\times10^{-2}$\,erg\,cm$^{-2}$ for the extrapolated range of $0.9-8,690$\,keV. We infer the isotropic-equivalent released energy of the first GRB episode to be $E_{\rm{iso}}^{\rm{bol}}=2.8_{-0.5}^{+0.8}\times10^{54}$\,erg in the $1-10,000$\,keV band (rest frame at $z=0.15$). The peak isotropic-equivalent luminosity in the $92-920$\,keV range (rest frame) was $L_{\rm{iso}}^{\rm{p}}=3.7_{-0.5}^{+2.5}\times10^{52}$\,erg s$^{-1}$ and the bolometric peak isotropic-equivalent luminosity was $L_{\rm{iso}}^{\rm{p,bol}}=8.4_{-1.5}^{+2.5}\times10^{52}$\,erg\,s$^{-1}$ (4\,s scale) in the $1-10,000$\,keV range (rest frame). The peak emitted energy is $E_p^\ast=E_p(1+z)=1120\pm470$\,keV. Our measurement of $L_{\rm{iso}}^{\rm{p,bol}}$ is consistent with the Yonetoku relation. It is possible that, due to the spectral evolution of this GRB and orientation of \textit{GRBAlpha} at the peak time, the true values of peak flux, fluence, $L_{\rm{iso}}$, and $E_{\rm{iso}}$ are even higher.}
   {}

   \maketitle

\section{Introduction}

On 2022 October 9 at 13:16:59.988 UT the \textit{Fermi} Gamma-ray Burst Monitor (GBM) detected the exceptionally bright long gamma-ray burst (GRB) GRB 221009A \citep{veres2022,lesage2022,lesage2023}. The burst was also observed by \textit{Fermi} Large Area Telescope (LAT) up to the energy of 100\,GeV \citep{pillera2022}. Potentially remarkable detections of over 5000 very high energy (VHE) photons with energies up to 18\,TeV were reported by the Large High Altitude Air Shower Observatory (LHAASO) \citep{huang2022} and a possible 251\,TeV photon was reported by Carpet-2  \citep{dzhappuev2022}, triggering the interest of the broader physics community. 

The burst was localised by Neil Gehrels \textit{Swift} Observatory’s Burst Alert Telescope \citep{dichiara2022} and followed up by Very Large Telescope's (VLT) X-shooter \citep{ugarte2022,Malesani2023}, which determined that it occurred at a redshift of 0.151 and it belongs to very near long GRBs \citep{Oates2023}.

It was detected also by a multitude of other instruments:
\textit{AGILE}/GRID \citep{Piano2022},
\textit{AGILE}/MCAL \citep{Ursi2022},
\textit{BepiColombo}/MGNS \citep{Kozyrev2022},
\textit{Insight}-HXMT \& \textit{SATech-01}/GECAM-C (HEBS) \citep{An2023}, 
\textit{INTEGRAL}/SPI-ACS \citep{Gotz2022},
Konus-\textit{WIND} \& \textit{SRG}/ART-XC \citep{Frederiks2023}, 
MAXI \& NICER \citep{Williams2023}, 
\textit{Solar Orbiter}/STIX \citep{Xiao2022},
\textit{STPSat-6}/SIRI-2 \citep{Mitchell2022}, and
\textit{XMM-Newton} \citep{Tiengo2023}.

This brightest ever recorded GRB \citep{Burns2023,OConnor2023} saturated many of the gamma-ray burst detectors in orbit, hampering the efforts to determine its peak luminosity. In this Letter, we present the measurement of the peak flux and peak isotropic-equivalent luminosity of this extraordinary transient with the \textit{GRBAlpha} nano-satellite.

\section{GRBAlpha}

\textit{GRBAlpha} \citep{pal2020} is a 1U CubeSat carrying a GRB detector as a technology demonstration for an envisioned future CubeSat constellation \citep{werner2018,meszaros2022}. It was launched on 2021 March 22 to a sun-synchronous polar orbit at an altitude of $\sim550$\,km and thus became the smallest astrophysical space observatory. About a third of the polar orbit is affected by high particle background and the duty cycle of the detector is around 67\%. \textit{GRBAlpha}’s detector consists of a $75\times 75\times 5$\,mm$^3$ CsI(Tl) scintillator read out by an array of Silicon PhotoMultipliers (SiPM), called multi-pixel photon counters (MPPCs), by Hamamatsu. The SiPM detectors are protected from proton damage by a 2.5\,mm thick lead (PbSb3 alloy) shield and their degradation is being monitored. The on-board data acquisition software stack is being continuously upgraded to increase the duty cycle and data downlink rate. The ground segment is also supported by the radio amateur community and it takes advantage of the Satellite Networked Open Ground Station (SatNOGS)\footnote{\url{https://satnogs.org}}. 

Following a commissioning phase, \textit{GRBAlpha} started collecting data, monitoring the particle and photon background environment on low-Earth polar orbit and detecting transients \citep[see][]{ripa2022a}. When the satellite operates continuously, on average it detects a transient every 5-6 days\footnote{The list of all \textit{GRBAlpha} detected transients is available here \url{https://monoceros.physics.muni.cz/hea/GRBAlpha/}}. \textit{GRBAlpha} detected GRB 221009A \citep{ripa2022b} when traversing the northern polar regions and during the peak brightness of the burst its detectors were not saturated, however, the measurement was influenced by pile-up.

The pile-up effect occurs by the chance coincidence of arrival and interaction of two or more gamma-ray photons in the detector's scintillator within the detector's inherent resolving time. In that case two or more events are registered as one event and the energies of the events within the same resolving time window are added \citep{Knoll2000}.

   \begin{figure*}
   \begin{center}
   \resizebox{0.49\textwidth}{!}{\includegraphics{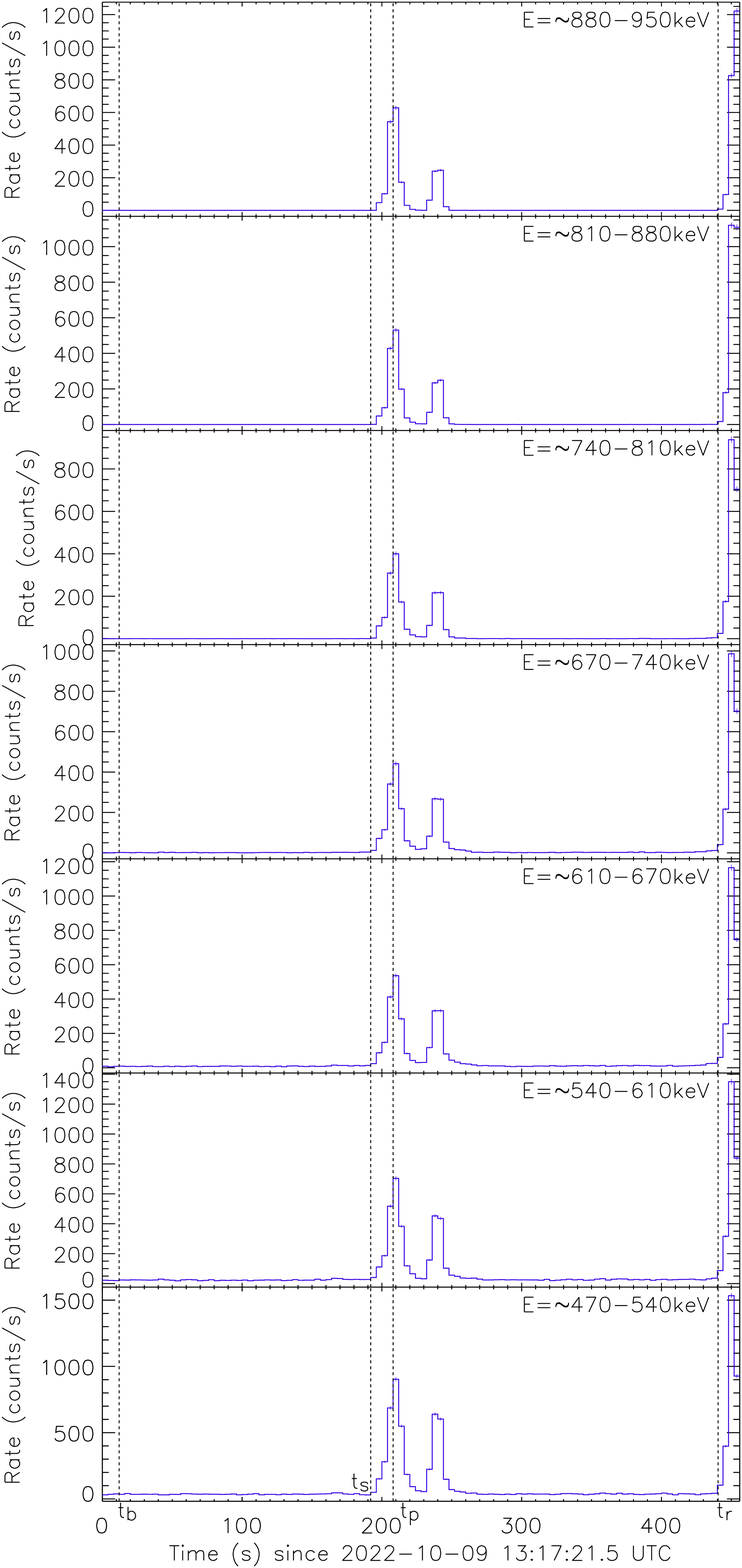}}
   \resizebox{0.49\textwidth}{!}{\includegraphics{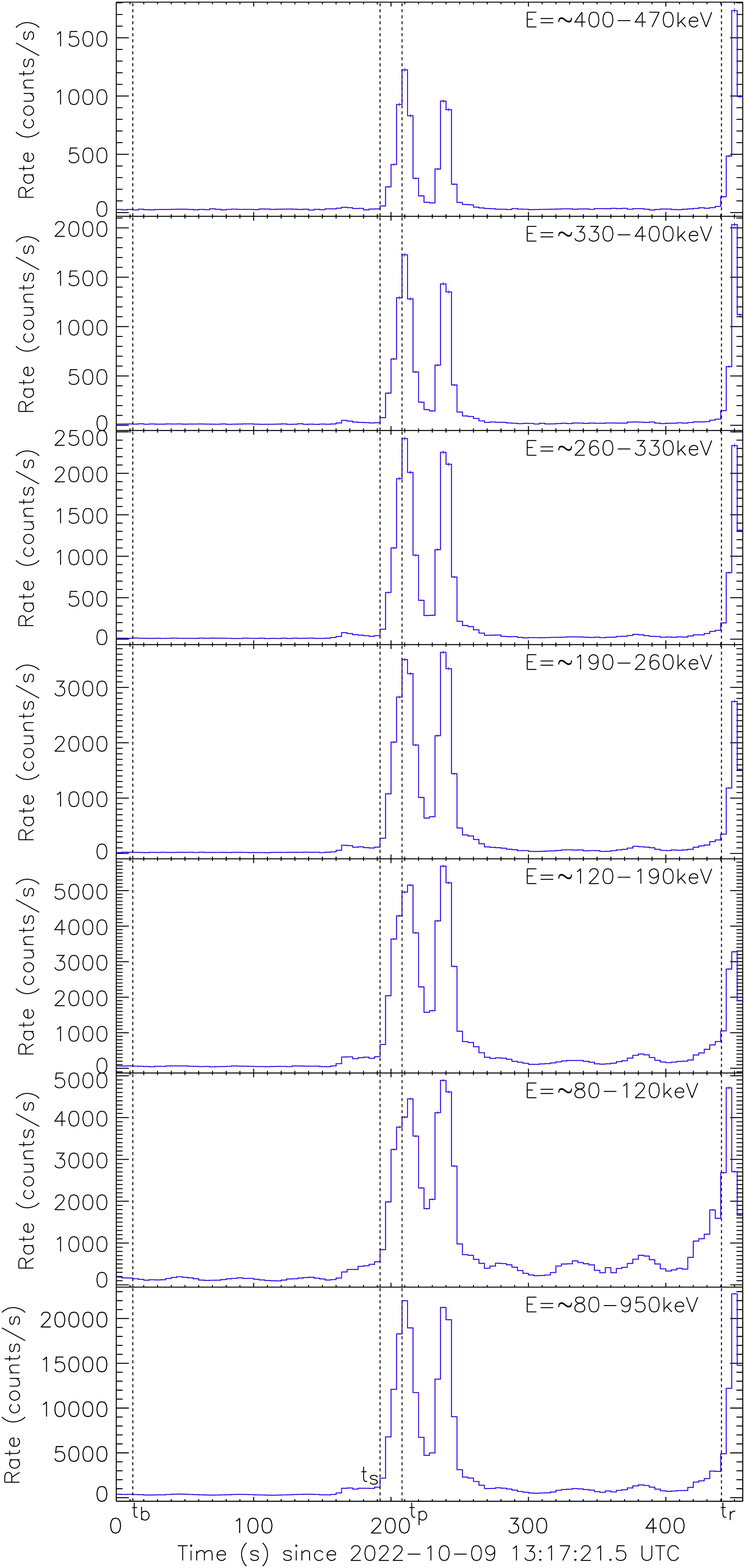}}
   \caption{The raw light curve of GRB 221009A as observed by \textit{GRBAlpha} in multiple energy bands with 4\,s cadence. The most bottom right panel shows the light curve in the whole sensitivity range. Times $t_b$, $t_s$ and $t_p$ mark the beginning of the used background interval, source spectral data and the peak time, respectively. About 30\,s after the peak at $t_p$ the GRB shows a second strong peak. Time $t_r$ marks the approximate moment when the satellite entered the outer Van Allen radiation belt, which resulted in the final part of the GRB prompt emission being flooded by particle background. The sharp increase just after $t_r$ is due to this background.}
   \label{fig:GRB_lc}
   \end{center}
   \end{figure*}

\section{Data analysis}

\textit{GRBAlpha} observed a peak count rate of $22\, 000$\,cnt\, s$^{-1}$ in the $\sim80-950$\,keV energy band at 2022-10-09 13:20:51.5 UTC. The duration of the GRB was $>250$\,s \citep{ripa2022b}. During the detection, the satellite was flying above the northern polar region with elevated background levels. The end part of the GRB was recorded while passing the outer Van Allen radiation belt, therefore we can only report the lower limit on the duration and we cannot determine the $T_{90}$ duration of the GRB. The \textit{GRBAlpha} data for this event are composed of binned light curves with 4\,s temporal resolution recorded in 16 energy bands. However, the three lowest energy bands are below the set on-board low-energy threshold limited by the noise peak of the MPPCs, making only 13 energy bands suitable for spectral analysis. The recorded raw count-rate curves in multiple energy bands are presented in Figure~\ref{fig:GRB_lc}. Note that the energy calibration was performed by using radionuclides in the laboratory and by observing activation lines in orbit.

\begin{figure}
   \centering
   \includegraphics[width=0.8\linewidth]{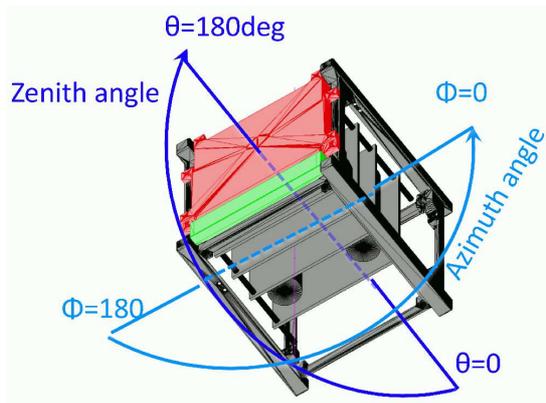}
      \caption{The mass model of \textit{GRBAlpha} used for generating detector response matrices with marked zenith and azimuth angles. The CsI(Tl) detector in its aluminium casing is shown in the red colour. The green colour marks the PbSb3 radiation shield. The grey colour marks the aluminium CubeSat platform, 4 stacks of PCBs and 2 LiFePO$_4$ batteries.}
         \label{fig:mass_model}
   \end{figure}

We perform spectral analysis to determine the peak flux of the GRB and its isotropic-equivalent peak luminosity. The mass model of \textit{GRBAlpha} used for generating detector response matrices employed in the spectral analysis is displayed in Figure~\ref{fig:mass_model}. It contains: the detector made of a $75\times75\times5$\,mm$^3$ CsI(Tl) scintillator in a 1.5\,mm thick Al casing; a 2.5\,mm thick PbSb3 radiation shield; a standard 1U CubeSat Al platform with four stacks of empty Printed Circuit Boards (PCBs) (glass epoxy FR4) and two LiFePO$_4$ batteries.

\begin{figure}
   \centering
   \includegraphics[height=0.9\linewidth, angle=270]{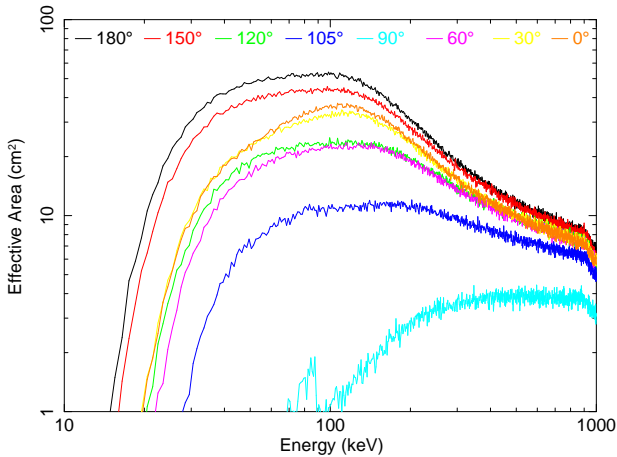}
      \caption{Total response efficiency (effective area) versus incident photon energy for few zenith angles $\Theta$ and constant azimuth angle $\Phi=270^\circ$.}
         \label{fig:effective_area}
   \end{figure}

Full Monte Carlo simulation based on GEometry ANd Tracking (Geant4) toolkit\footnote{\url{https://geant4.web.cern.ch/}} \citep{geant4} was carried out to simulate incident photons with a flat energy distribution in the $5-1,000$\,keV band covering the entire satellite structure. The deposited energy as a function of input photon energy was calculated scanning more than 200 incident angles $\Theta$ (zenith) and $\Phi$ (azimuth). The effective area versus incident photon energy for a few zenith angles $\Theta$ and a constant azimuth angle $\Phi=270^\circ$ is shown in Figure~\ref{fig:effective_area}. The drop over 900\,keV is an artefact because we simulated photon energies only up to 1,000\,keV and the photoabsorption peak (with broadening of the energy resolution) of such high energy photons is not accounted in this response. Therefore, in the spectral analysis, we conservatively only use data up to $\sim810$\,keV.

\begin{figure}
   \centering
   \includegraphics[width=\linewidth]{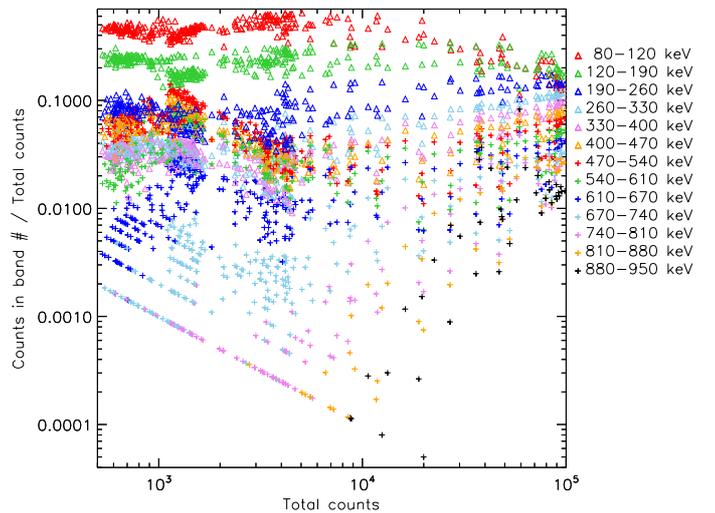}
      \caption{The fraction of the detected counts in each spectral band vs the total counts over the whole energy range for GRB 221009A and for background while passing the north polar region. The counts were detected in 4\,s time bins. The figure shows that above $\sim10,000$\,cnt/bin = 2,500\,cnt/s the fraction of counts in low-energy bands decreases while the fraction of counts in high-energy bands increases which can indicate the pile-up effect.}
         \label{fig:counts_ratio}
   \end{figure}

To check for pile-up, we plot the fraction of the detected counts in each spectral band versus the total counts over the whole energy range for GRB 221009A (see Figure~\ref{fig:counts_ratio}). The figure shows that above $\sim10,000$\,cnt/bin = 2,500\,cnt\,s$^{-1}$ the fraction of counts in the lower-energy bands decreases while the fraction of counts in the higher-energy bands increases, which may indicate the presence of pile-up. We note that below $\sim10,000$\,cnt/bin the decrease of the rates at high energies as a function of total counts in each bin is due to very low count numbers and the diagonal lines correspond to 1, 2, $3\dots$ cnt/bin. At the peak of this transient \textit{GRBAlpha} measured a count rate of $m=22,000$\,cnt\,s$^{-1}$ in the full band. The shaping parameter of the analogue electronics is $\tau_1\approx10\,\mu$s. During the signal digitisation the pulse processing is done within the resolution time of $\tau_2=15\,\mu$s. We can calculate the probability that a pile-up appears as \citep{Knoll2000}:
\begin{equation}
P=1-e^{-n\tau_2},
\end{equation}
where $n$ is the true event rate which can be calculated from the detected count rate $m$ as:
\begin{equation}
n=-\frac{\rm{ln}(1-m\tau)}{\tau},
\end{equation}
where for $m=22,000$\,cnt\,s$^{-1}$ we obtain $n=26,700$\,event\,s$^{-1}$.
For the observed peak count rate we obtain the pile-up probability of $\sim33$\% whereas the probability drops down to $\sim3.7$\% for a count rate of $m=2,500$\,cnt\,s$^{-1}$. Therefore, the observed spectrum at the peak time is affected by pile-up and in the following spectral analysis we avoid fitting the peak spectrum (which results in a flatter power-law fit) and rather fit the GRB spectrum before the peak time when the pile-up is negligible.

\begin{figure}
   \centering
   \includegraphics[height=0.9\linewidth, angle=270]{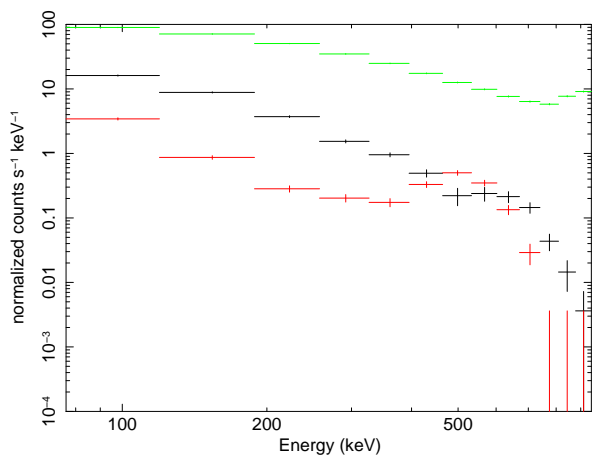}
      \caption{Black: the detected count spectrum of GRB 221009A at time $t_s$ minus the background at time $t_b$. Red: the background at time $t_b$. Green: the detected count spectrum of GRB 221009A at the peak time $t_p =$ 2022-10-09 13:20:49.5 UT, minus the background at time $t_b$, showing the spectral hardening at highest energies present around the GRB peak.}
         \label{fig:grb_bkg_spectra}
   \end{figure}

The spectral fitting was performed by the X-ray Spectral Fitting Package XSPEC\footnote{\url{https://heasarc.gsfc.nasa.gov/xanadu/xspec/}} \citep{Arnaud1996}. The source spectrum was fitted from time $t_s =$ 13:20:33.5 UTC to $t_s+4$\,s and the background spectrum was taken from time $t_b =$ 13:17:33.5 UTC to $t_b+4$\,s.
By applying this background, we obtained a background-subtracted spectrum without any substantial bump or dip around 500\,keV where the \textit{GRBAlpha} background manifest a bump. The average background rate in the whole sensitivity range in this part of orbit is 340\,cnt\,s$^{-1}$ and the rate in the chosen 4\,s background bin is 345\,cnt\,s$^{-1}$.
We set a systematic error of 2\% to be added when evaluating $\chi^2$. The peak time in the measured light curve in the whole sensitivity range of $\sim 80-950$\,keV was measured to be from $t_p =$ 13:20:49.5 UTC to $t_p+4$\,s. Figure~\ref{fig:grb_bkg_spectra} displays the detected count spectra at the time bins $t_s$, $t_b$ and $t_p$. At an approximate time $t_r =$ $\sim$13:24:42.0 UTC the \textit{GRBAlpha} nano-satellite entered the outer Van Allen radiation belt and the signal from the GRB became flooded by particle background. These times are marked in Figure~\ref{fig:GRB_lc}.

\textit{GRBAlpha} does not have an active attitude control system and the attitude knowledge was not being recorded at the time of GRB 221009A. The background variation had a periodicity of $\sim47$\,s when the satellite was near the north pole. From the communication with the satellite on the same day, we observed in the radio waterfall a periodicity of $\sim23$\,s in the received radio power. The Ultra High Frequency (UHF) antenna on \textit{GRBAlpha} is a half-wave dipole and thus a periodicity in the waterfall corresponds to half of the rotation or precession period of the satellite. Hence the satellite completed more then one revolution during the duration of the GRB. Therefore, we performed the spectral analysis for many detector response matrices simulated for various incident angles $\Theta$ and $\Phi$. We fitted the measured spectra each time in order to obtain the best-fit $\chi^2$ map and to obtain the most likely direction of the GRB with respect to the satellite's coordinate system.

We performed the spectral fitting with the power-law model, power-law with an exponential cutoff (CPL), and with the Band function. The lowest reduced $\chi^2$ was obtained for the CPL model, as described in the following section. The reduced $\chi^2$ maps show a similar trend, therefore, in the next section, we report results obtained with the CPL model only.

After we determined the most likely incidence angle of the GRB with respect to the detector, we calculated the flux and the isotropic-equivalent luminosity of the GRB using the best-fit spectral model. Scaling-up the flux and the luminosity obtained from the spectral fit at time $t_s$ allows us to determine the flux and the luminosity at the peak time $t_p$. The scaling factor is approximately the ratio of the total detected counts at the time bin $t_p$ over the total detected counts at the time bin $t_s$. The results are described in detail in the following section.

\section{Results}

\begin{figure}
   \centering
   \includegraphics[height=0.9\linewidth, angle=270]{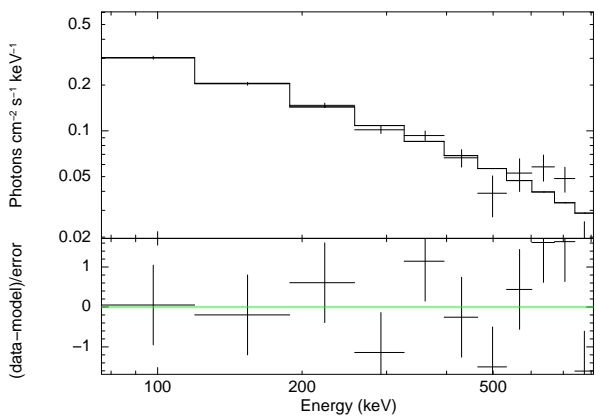}
      \caption{The spectrum of the GRB 221009A best fit by the CPL model fitted from $t_s =$ 2022-10-09 13:20:33.5 UTC to $t_s+4$\,s in the range of $76-809$\,keV for $\Theta=180^\circ$ and $\Phi=0^\circ$.}
         \label{fig:GRB_spectrum_ts}
   \end{figure}

The best fit for the source spectrum at the time bin $t_s$ and after subtraction of the background at the time bin $t_b$ was obtained for the on-axis direction $\Theta=180^\circ$ and $\Phi=0^\circ$ by the CPL model:
\begin{equation}
N(E)=A\left ( \frac{E}{\rm{1keV}} \right )^{-\alpha}\rm{exp}\left (-\frac{E}{E_{0}} \right ),
\end{equation}
see Figure~\ref{fig:GRB_spectrum_ts}. The fit was performed in the range spanning $76-809$\,keV giving $\chi^2/\rm{DoF}=13.4/8=1.68$. The best-fit parameters are: the photon index $\alpha=0.7\pm0.1$; the rolloff energy $E_0=750_{-200}^{+410}$\,keV and the normalisation $A=8_{-4}^{+6}$\,ph\,keV$^{-1}$cm$^{-2}$s$^{-1}$ at 1\,keV. The relatively large uncertainty of the normalisation is due to its correlation with the photon index alpha and the rolloff energy. The 68\% confidence interval (CI) parameter uncertainties were calculated by XSPEC from the fit covariance matrix. The peak energy is $E_p=E_0(2-\alpha)=980\pm410$\,keV.

The flux in the energy range of $80-800$\,keV in the observer frame for the best fit model for the same time interval $t_s$ and the same $\Theta$ and $\Phi$ angles was derived to be $F_{\rm{ph}}^{\rm{s}}=68\pm1$\,ph\,cm$^{-2}$s$^{-1}$ or $F_{\rm{erg}}^{\rm{s}}=3.3\pm0.1\times10^{-5}$\,erg\,cm$^{-2}$s$^{-1}$ (68\% CI). For the isotropic-equivalent luminosity in the energy range of $92-920$\,keV in the rest frame for the redshift of $z = 0.151$ \citep{ugarte2022,Malesani2023} and cosmological parameter $\Omega_{\Lambda}=0.685$, flat universe and Hubble constant of $H_{0}=67.4$\,km\,s$^{-1}$\,Mpc$^{-1}$ \citep{Planck_Collaboration2020} it was obtained $L_{\rm{iso}}^{\rm{s}}=2.2_{-0.0}^{+0.1}\times10^{51}$\,erg\,s$^{-1}$ (68\% CI). The uncertainties in flux and luminosity were derived from Markov Chain Monte Carlo \citep{Hastings1970} with following parameters: proposal Gaussian fit; re-scale the covariance matrix used in the proposal distribution by the factor of 0.125; chain length of 50,000; discard the first 1,000 steps prior to storing the chain and Goodman-Weare chain type with 16 walkers.

The approximate scaling factor $f=(C_p-C_b) / (C_s-C_b)$ which is the ratio between the total detected counts $C_p$ at the peak time bin $t_p$ minus mean background counts $C_b=1,360$\,cnt/bin $=340$\,cnt\,s$^{-1}$ and the total detected counts $C_s$ at the time bin $t_s$ minus the mean background counts which gives $f=(87,950-1,360) / (8,689-1,360) = 11.8$. We scale the flux and the luminosity obtained from the spectral fit at time $t_s$ by this factor and by the pile-up correction factor of 1.21 in order to determine the flux and the luminosity at the peak time $t_p$. By applying this scaling to the flux and the isotropic-equivalent luminosity derived above for the time bin $t_s$ we obtain for the peak flux $F_{\rm{ph}}^{\rm{p}}=970\pm20$\,ph\,cm$^{-2}$s$^{-1}$ or $F_{\rm{erg}}^{\rm{p}}=4.6\pm0.2\times10^{-4}$\,erg\,cm$^{-2}$s$^{-1}$ and for the peak luminosity $L_{\rm{iso}}^{\rm{p}}=3.1\pm0.1\times10^{52}$\,erg\,s$^{-1}$.

However, so far we did not consider the systematic uncertainty due to the uncertain attitude of the satellite (detector). Although, we derived the most likely orientation of the satellite with respect to the GRB from the best-fit $\chi^2$ map, the difference in reduced $\chi^2$ near the on-axis direction ($\Theta=180^\circ$) and near the backside direction ($\Theta=0^\circ$) is relatively low implying that the on-axis direction is not guaranteed. Therefore, we also calculated the flux and the luminosity using a more conservative approach.

\begin{figure}
   \centering
   \includegraphics[width=0.9\linewidth]{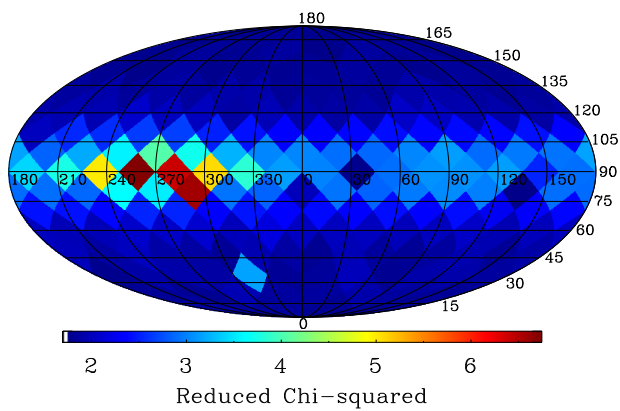}
      \caption{The reduced $\chi^2$ map of the best fits with CPL model for different angles $\Theta$ (vertical) and $\Phi$ (horizontal) in degrees.}
         \label{fig:chi2_map}
   \end{figure}

We took the distribution of fluxes and luminosities obtained by fitting the spectra by the CPL model at the same time bin $t_s$, with the same background time bin $t_b$ as described above, and used 192 different directions $\Theta$ and $\Phi$ which isotropically sample the sphere around the satellite following the Hierarchical Equal Area isoLatitude Pixelization (HEALPix)\footnote{\url{https://healpix.sourceforge.io}} tessellation \citep{Gorski2005}. The best-fit $\chi^2$ was obtained for each direction. The reduced $\chi^2_{\rm{r}}$ map is shown in Figure~\ref{fig:chi2_map}. We removed 6 cases which have reduced $\chi^2_{\rm{r}}>4$ because these we consider as not acceptable fits. They correspond to the direction towards the MPPCs' lead shield. The resulting distribution contains 186 fluxes (luminosities). The median flux in the $80-800$\,keV range (observer frame) and the isotropic-equivalent luminosity in the $92-920$\,keV range (rest frame) at the same time bin $t_s$ are:
$F_{\rm{ph}}^{\rm{s}}=94_{-17}^{+85}$\,ph\,cm$^{-2}$s$^{-1}$ or 
$F_{\rm{erg}}^{\rm{s}}=4.0_{-0.5}^{+2.6}\times10^{-5}$\,erg\,cm$^{-2}$s$^{-1}$ and
$L_{\rm{iso}}^{\rm{s}}=2.6_{-0.3}^{+1.7}\times10^{51}$\,erg\, s$^{-1}$.

 By applying the scaling factor $f=(C_p-C_b) / (C_s-C_b) = 11.8$ and the pile-up correction factor of 1.21 we obtain for the peak flux
 $F_{\rm{ph}}^{\rm{p}}=1300_{-200}^{+1200}$\,ph\,cm$^{-2}$s$^{-1}$ or
 $F_{\rm{erg}}^{\rm{p}}=5.7_{-0.7}^{+3.7}\times10^{-4}$\,erg\,cm$^{-2}$s$^{-1}$ and 
for the isotropic-equivalent peak luminosity
$L_{\rm{iso}}^{\rm{p}}=3.7_{-0.5}^{+2.5}\times10^{52}$\,erg\,s$^{-1}$. When we repeat this analysis for the fluxes integrated over larger extrapolated energy range of $0.9-8,690$\,keV, we obtain a peak flux $F_{\rm{erg}}^{\rm{p,bol}}=1.3_{-0.2}^{+0.4}\times10^{-3}$\,erg\,cm$^{-2}$s$^{-1}$ and a bolometric isotropic-equivalent peak luminosity $L_{\rm{iso}}^{\rm{p,bol}}=8.4_{-1.5}^{+2.5}\times10^{52}$\,erg\,s$^{-1}$ (4\,s scale) in $1-10,000$\,keV range (rest frame).

Results for the on-axis direction and those obtained by varying different incident directions are summarised in Table~\ref{tab:results}.

Having the peak flux $F_{\rm{erg}}^{\rm{p}}$ we can calculate the approximate fluence $S$ of the GRB as:
\begin{equation}
S=\tfrac{C-kC_b}{C_p-C_b}F_{\rm{erg}}^{\rm{p}}\Delta t,
\end{equation}
where $C$ are total detected counts during the GRB, $C_p$ are detected counts at the peak time bin $t_p$ of duration $\Delta t=4$\,s, $C_b$ are mean counts per bin and $k$ is the number of bins over which we calculate the fluence. We obtained the fluence $S=2.2_{-0.3}^{+1.4}\times10^{-2}$\,erg\,cm$^{-2}$ in the $80-800$\,keV range of the first GRB episode lasting 300\,s which was observable by \textit{GRBAlpha} from time 2022-10-09 13:19:29.5 to 13:24:29.5 UTC. In the same way but using the wider extrapolated energy range of $0.9-8,690$\,keV we obtain the fluence $S^{\rm{bol}}=4.9_{-0.5}^{+0.8}\times10^{-2}$\,erg\,cm$^{-2}$ and the
isotropic-equivalent released energy $E_{\rm{iso}}^{\rm{bol}}=2.8_{-0.5}^{+0.8}\times10^{54}$\,erg in the $1-10,000$\,keV range (rest frame).

\begin{table*}
\caption{The summary of obtained fluxes and luminosities for the on-axis direction ($\Theta=180^\circ$, $\Phi=0^\circ$), the median values obtained by varying different incident angles and the values for the edge-on direction ($\Theta=90^\circ$, $\Phi=101^\circ$). The values marked with the superscript ``s'' and ``p'' were derived for the time bin $t_s$ and the peak time $t_p$, respectively. Fluxes are for the $80-800$\,keV range (observer frame) and luminosities are for the $92-920$\,keV range (rest frame).}
\label{tab:results}
\centering
\begin{tabular}{ccccccc}
\hline\hline \\[-10pt]
Method & $F_{\rm{ph}}^{\rm{s}}$ & $F_{\rm{erg}}^{\rm{s}}$ & $L_{\rm{iso}}^{\rm{s}}$ & $F_{\rm{ph}}^{\rm{p}}$ & $F_{\rm{erg}}^{\rm{p}}$ & $L_{\rm{iso}}^{\rm{p}}$ \\
 & (ph\,cm$^{-2}$s$^{-1}$) & ($10^{-5}$\,erg\,cm$^{-2}$s$^{-1}$) & ($10^{51}$\,erg\, s$^{-1}$) & (ph\,cm$^{-2}$s$^{-1}$) & ($10^{-4}$\,erg\,cm$^{-2}$s$^{-1}$) & ($10^{52}$\,erg\, s$^{-1}$) \\
\hline \\[-9pt]
On-axis & $68\pm1$   & $3.3\pm0.1$ & $2.2_{-0.0}^{+0.1}$ & $970\pm20$ & $4.6\pm0.2$ & $3.1\pm0.1$ \\[2pt]
Median  & $94_{-17}^{+85}$ & $4.0_{-0.5}^{+2.6}$ & $2.6_{-0.3}^{+1.7}$ & $1300_{-200}^{+1200}$ & $5.7_{-0.7}^{+3.7}$ & $3.7_{-0.5}^{+2.5}$ \\[2pt]
Edge-on & $390\pm10$ & $13.7_{-0.1}^{+0.4}$ & $9.1_{-0.1}^{+0.3}$ & $5600\pm100$ & $19.6_{-0.1}^{+0.6}$ & $12.9_{-0.2}^{+0.4}$ \\[2pt]
\hline
\end{tabular}
\end{table*}

\section{Discussion}

The highest peak flux and fluence of all GRBs recorded with {\it CGRO}/BATSE \citep{1999ApJS..122..465P} and {\it Fermi}/GBM \citep{2020ApJ...893...46V} are 496\,ph\,cm$^{-2}$\,s$^{-1}$ (50--300 keV) and $8.1\times10^{-4}$\,erg\,cm$^{-2}$  (50--300 keV) for GRB 130427A. The measured peak flux and fluence of GRB 221009A are $1020_{-160}^{+950}$\,ph\,cm$^{-2}$\,s$^{-1}$ and $8.3_{-1.1}^{+5.3}\times10^{-3}$\,erg\,cm$^{-2}$ in the 50--300\,keV band. The duration of GRB 221009A is longer than 100\,s, compared to the 10--20\,s for GRB 130427A, giving a relatively larger fluence ratio.
This indicates that GRB 221009A is indeed exceptionally bright.
While the size of the CsI scintillator on \textit{GRBAlpha} is not small ($\sim55$\,cm$^2$ face-on), the low-energy threshold of the detector is relatively high at around 80\,keV due to the radiation damage of the SiPM detectors. This allowed \textit{GRBAlpha} to avoid saturation of the count rate in the digital processing, allowing us to trace the light curve. Note that the brightest gamma-ray flux from a celestial object was observed during the giant flare from the magnetar SGR 1806-20 in 2004 and its peak flux was around 10--20\,erg\,s$^{-1}$ in the 20--10,000\,keV band \citep{2005astro.ph..2541M,2005Natur.434.1110T}, which is 4--5 orders of magnitude higher than that of GRB 221009A.

We note that this exceptionally bright GRB has one of the highest isotropic-equivalent peak luminosities ever reported \citep{Burns2023}. GRBs with higher peak luminosities were observed in the past, but at significantly larger redshifts, see e.g. \citet{Yonetoku2010}. In their Figure~3 the authors provide the correlation between the rest-frame spectral peak energy and 1-second peak luminosity of short and long GRBs. Our measurement of $L_{\rm{iso}}^{\rm{p,bol}}=8.4_{-1.5}^{+2.5}\times10^{52}$\,erg\,s$^{-1}$ and 
$E_p^\ast=E_p(1+z)=1120\pm470$\,keV is consistent with this relation within the systematic errors.

In Figure~\ref{fig:GRB_lc} one can see that at low energies, mainly at $80-120$\,keV, the background shows a wavy pattern. This is most likely due to the periodic motion of \textit{GRBAlpha}, exposing a different cross-section of the detector to the Cosmic X-ray Background (CXB) and the secondary gamma-rays induced in the Earth's atmosphere by cosmic rays \citep{Galgoczi2021}. On the same day that we detected this GRB, we observed a periodicity also in the received radio power in the waterfalls, while downloading data from the satellite, revealing that the satellite completed more then one revolution during the duration of the GRB. Therefore, change of the satellite's attitude between times $t_s$ and $t_p$ can result in a systematic error of the reported peak luminosity and total emitted energy. As shown in Table~\ref{tab:results}, the luminosity derived for the edge-on direction is $3.5$ times higher than the median value. We cannot exclude that the detector on board of \textit{GRBAlpha} was oriented edge-on towards the GRB at the time of the peak and therefore, we cannot exclude that the obtained median peak flux and luminosity are underestimated.

The time-resolved spectral analysis of GRB 221009A by Konus-\textit{WIND} \citep{Frederiks2023} reveals spectral evolution during the burst with higher photon index $\alpha$ and peak energy $E_{\rm{p}}$ around the peak time reaching $E_{\rm{p}}\approx3$\,MeV. Hence our method based on scaling the flux from the spectrum at time $t_s$ to peak time $t_p$ may underestimate the resultant peak flux and fluence. Therefore, it is possible that the true peak flux and fluence are higher.

The highest energy bands around $800-950$\,keV as shown in Figure~\ref{fig:GRB_lc} have increased number of counts during the GRB peak compared to lower energy bands around $700-800$\,keV. This is also evident in the raw count spectrum at the peak time $t_p$ presented in Figure~\ref{fig:grb_bkg_spectra}. This could be due to an instrumental effect because of the high incident photon rate.

\section{Conclusions}

The conclusions can be summarised in these main points:

\begin{itemize}
    \item \textit{GRBAlpha}, a low-cost nano-satellite on a low-Earth orbit, has detected the exceptionally bright long gamma-ray burst GRB 221009A without saturation, but effected by pile-up, providing light curves of the prompt emission in 13 energy bands from 80\,keV to 950\,keV.
    \item The peak flux in the $80-800$\,keV range (observer frame) was measured to be $F_{\rm{ph}}^{\rm{p}}=1300_{-200}^{+1200}$\,ph\,cm$^{-2}$s$^{-1}$ or $F_{\rm{erg}}^{\rm{p}}=5.7_{-0.7}^{+3.7}\times10^{-4}$\,erg\,cm$^{-2}$s$^{-1}$ and for the wider extrapolated energy range of $0.9-8,690$\,keV, we obtained $F_{\rm{erg}}^{\rm{p,bol}}=1.3_{-0.2}^{+0.4}\times10^{-3}$\,erg\,cm$^{-2}$s$^{-1}$.
    \item The peak isotropic-equivalent luminosity in the $92-920$\,keV range (rest frame) was $L_{\rm{iso}}^{\rm{p}}=3.7_{-0.5}^{+2.5}\times10^{52}$\,erg s$^{-1}$ and the bolometric peak isotropic-equivalent luminosity in the $1-10,000$\,keV range (rest frame) was $L_{\rm{iso}}^{\rm{p,bol}}=8.4_{-1.5}^{+2.5}\times10^{52}$\,erg\,s$^{-1}$ (4\,s scale). Our measurement of $L_{\rm{iso}}^{\rm{p,bol}}$ is consistent with the Yonetoku relation between the rest-frame spectral peak energy and the peak luminosity of GRBs.
    \item The fluence in the $80-800$\,keV range of the first GRB episode lasting 300\,s which was observable by \textit{GRBAlpha} from 2022-10-09 13:19:29.5 to 13:24:29.5 UTC was measured to be $S=2.2_{-0.3}^{+1.4}\times10^{-2}$\,erg\,cm$^{-2}$ and $S^{\rm{bol}}=4.9_{-0.5}^{+0.8}\times10^{-2}$\,erg\,cm$^{-2}$ for the extrapolated range of $0.9-8,690$\,keV. We infer from this extrapolated fluence that the isotropic-equivalent released energy of the first GRB episode was $E_{\rm{iso}}^{\rm{bol}}=2.8_{-0.5}^{+0.8}\times10^{54}$\,erg in the $1-10,000$\,keV band (rest frame).
    \item It is possible that due to the spectral evolution during the peaks of this GRB, as reported by Konus-\textit{WIND} \citep{Frederiks2023}, the true peak flux, fluence, $L_{\rm{iso}}$ and $E_{\rm{iso}}$ can be higher than our obtained values. We cannot exclude that the \textit{GRBAlpha}'s detector was oriented near edge-on with respect to the GRB 221009A at the peak time which would also mean that the true values can be higher than our estimations.
\end{itemize}
 
\begin{acknowledgements}
We are thankful to Peter Veres and Stephen Lesage for discussions regarding the detection of this GRB by \textit{Fermi}/GBM and other instruments and to Tom\'a\v{s} Pl\v{s}ek for discussions about MCMC and XSPEC.
We are also thankful to Nozomu Kogiso, Maumu Yoneyama, Mizuki Moritaki, Tatsuya Kano and James W. Cutler for providing their SatNOGS stations for collecting \textit{GRBAlpha}'s telemetry.
We acknowledge support by the  KEP-7/2018 and KEP2/2020 grants of the Hungarian Academy of Sciences and SA-40/2021 of the E\"otv\"os Lor\'and Research Network for satellite components and payload developments and the grant IF-7/2020 for providing the financial support for ground infrastructure. This research has been supported by the European Union’s Horizon 2020 programme under the AHEAD2020 project (grant agreement n. 871158) and by the MUNI Award for Science and Humanities funded by the Grant Agency of Masaryk University. GD is supported by the Ghent University Special Research Funds (BOF) project BOF/STA/202009/040 and the Fonds Wetenschappelijk Onderzoek (FWO) iBOF project BOF20/IBF/124. This work was supported by the Internal Grant Agency of Brno University of Technology, project no. FEKT-S-20-6526. This research has been also supported by JSPS and HAS under Japan-Hungary Research Cooperative Program, JSPS KAKENHI Grant Number 17H06362, 19H01908, and 21KK0051. We are grateful to the Ministry of Education, Science, Research and Sport of the Slovak Republic for the support of the GRBAlpha mission.
\end{acknowledgements}

\bibliographystyle{aa} 
\bibliography{references.bib} 

\begin{thebibliography}{39}
\expandafter\ifx\csname natexlab\endcsname\relax\def\natexlab#1{#1}\fi

\bibitem[{{Allison} {et~al.}(2016){Allison}, {Amako}, {et~al.}}]{geant4}
{Allison}, J., {Amako}, K., {et~al.} 2016, Nuclear Instruments and Methods in
  Physics Research Section A: Accelerators, Spectrometers, Detectors and
  Associated Equipment, 835, 186

\bibitem[{{An} {et~al.}(2023){An}, {Antier}, {Bi}, {Bu}, {Cai}, {Cao},
  {Camisasca}, {Chang}, {Chen}, {Chen}, {Chen}, {Chen}, {Chen}, {Chen}, {Chen},
  {Coughlin}, {Cui}, {Dai}, {Hussenot-Desenonges}, {Du}, {Du}, {Du}, {Fan},
  {Frontera}, {Gao}, {Gao}, {Ge}, {Gong}, {Gu}, {Guan}, {Guo}, {Guo},
  {Guidorzi}, {Han}, {He}, {He}, {Hou}, {Huang}, {Huo}, {Ji}, {Jia}, {Jiang},
  {Kann}, {Klotz}, {Kong}, {Lan}, {Li}, {Li}, {Li}, {Li}, {Li}, {Li}, {Li},
  {Li}, {Li}, {Li}, {Li}, {Li}, {Li}, {Liang}, {Liang}, {Liao}, {Lin}, {Liu},
  {Liu}, {Liu}, {Liu}, {Liu}, {Liu}, {Liu}, {Lu}, {Lu}, {Lu}, {Luo}, {Luo},
  {Ma}, {Ma}, {Ma}, {Ma}, {Maccary}, {Mao}, {Meng}, {Nie}, {Orlandini}, {Ou},
  {Peng}, {Peng}, {Qiao}, {Qu}, {Ren}, {Shi}, {Shi}, {Song}, {Song}, {Su},
  {Sun}, {Sun}, {Sun}, {Tan}, {Tan}, {Tao}, {Tuo}, {Turpin}, {Wang}, {Wang},
  {Wang}, {Wang}, {Wang}, {Wang}, {Wang}, {Wang}, {Wang}, {Wang}, {Wang},
  {Wang}, {Wang}, {Wang}, {Wen}, {Wu}, {Wu}, {Wu}, {Xiao}, {Xiao}, {Xiao},
  {Xie}, {Xiong}, {Xiong}, {Xu}, {Xu}, {Xu}, {Xu}, {Xu}, {Xu}, {Xue}, {Yang},
  {Yang}, {Yang}, {Ye}, {Yi}, {Yi}, {Yin}, {You}, {Yu}, {Yu}, {Yu}, {Zeng},
  {Zhang}, {Zhang}, {Zhang}, {Zhang}, {Zhang}, {Zhang}, {Zhang}, {Zhang},
  {Zhang}, {Zhang}, {Zhang}, {Zhang}, {Zhang}, {Zhang}, {Zhang}, {Zhang},
  {Zhang}, {Zhang}, {Zhang}, {Zhao}, {Zhao}, {Zhao}, {Zhao}, {Zhao}, {Zhao},
  {Zhao}, {Zhao}, {Zheng}, {Zheng}, {Zhou}, {Zhou}, \& {Zhu}}]{An2023}
{An}, Z.-H., {Antier}, S., {Bi}, X.-Z., {et~al.} 2023, arXiv e-prints,
  arXiv:2303.01203

\bibitem[{{Arnaud}(1996)}]{Arnaud1996}
{Arnaud}, K.~A. 1996, in Astronomical Society of the Pacific Conference Series,
  Vol. 101, Astronomical Data Analysis Software and Systems V, ed. G.~H.
  {Jacoby} \& J.~{Barnes}, 17

\bibitem[{{Burns} {et~al.}(2023){Burns}, {Svinkin}, {Fenimore}, {Feliciano
  Ag{\"u}{\'\i} Fern{\'a}ndez}, {Frederiks}, {Kann}, {Hamburg}, {Lesage},
  {Temiraev}, {Tsvetkova}, {Bissaldi}, {Briggs}, {Fletcher}, {Goldstein},
  {Hui}, {Hristov}, {Kocevski}, {Lysenko}, {Mailyan}, {Racusin}, {Ridnaia},
  {Roberts}, {Ulanov}, {Veres}, {Wilson-Hodge}, \& {Wood}}]{Burns2023}
{Burns}, E., {Svinkin}, D., {Fenimore}, E., {et~al.} 2023, arXiv e-prints,
  arXiv:2302.14037

\bibitem[{{de Ugarte Postigo} {et~al.}(2022){de Ugarte Postigo}, {Izzo},
  {Pugliese}, {Xu}, {Schneider}, {Fynbo}, {Tanvir}, {Malesani}, {Saccardi},
  {Kann}, {Wiersema}, {Gompertz}, {Thoene}, {Levan}, \& {Stargate
  Collaboration}}]{ugarte2022}
{de Ugarte Postigo}, A., {Izzo}, L., {Pugliese}, G., {et~al.} 2022, GRB
  Coordinates Network, 32648, 1

\bibitem[{{Dichiara} {et~al.}(2022){Dichiara}, {Gropp}, {Kennea}, {Kuin},
  {Lien}, {Marshall}, {Tohuvavohu}, \& {Williams}}]{dichiara2022}
{Dichiara}, S., {Gropp}, J.~D., {Kennea}, J.~A., {et~al.} 2022, The
  Astronomer's Telegram, 15650, 1

\bibitem[{{Dzhappuev} {et~al.}(2022){Dzhappuev}, {Afashokov}, {Dzaparova},
  {Dzhatdoev}, {Gorbacheva}, {Karpikov}, {Khadzhiev}, {Klimenko}, {Kudzhaev},
  {Kurenya}, {Lidvansky}, {Mikhailova}, {Petkov}, {Podlesnyi}, {Pozdnukhov},
  {Romanenko}, {Rubtsov}, {Troitsky}, {Unatlokov}, {Vaiman}, {Yanin}, \&
  {Zhuravleva}}]{dzhappuev2022}
{Dzhappuev}, D.~D., {Afashokov}, Y.~Z., {Dzaparova}, I.~M., {et~al.} 2022, The
  Astronomer's Telegram, 15669, 1

\bibitem[{{Frederiks} {et~al.}(2023){Frederiks}, {Svinkin}, {Lysenko},
  {Molkov}, {Tsvetkova}, {Ulanov}, {Ridnaia}, {Lutovinov}, {Lapshov},
  {Tkachenko}, \& {Levin}}]{Frederiks2023}
{Frederiks}, D., {Svinkin}, D., {Lysenko}, A.~L., {et~al.} 2023, arXiv
  e-prints, arXiv:2302.13383

\bibitem[{{Galg{\'o}czi} {et~al.}(2021){Galg{\'o}czi}, {{\v{R}}{\'\i}pa},
  {Campana}, {Werner}, {P{\'a}l}, {Ohno}, {M{\'e}sz{\'a}ros}, {Mizuno},
  {Tarcai}, {Torigoe}, {Uchida}, {Fukazawa}, {Takahashi}, {Nakazawa}, {Hirade},
  {Hirose}, {Hisadomi}, {Enoto}, {Odaka}, {Ichinohe}, {Frei}, \&
  {Kiss}}]{Galgoczi2021}
{Galg{\'o}czi}, G., {{\v{R}}{\'\i}pa}, J., {Campana}, R., {et~al.} 2021,
  Journal of Astronomical Telescopes, Instruments, and Systems, 7, 028004

\bibitem[{{G{\'o}rski} {et~al.}(2005){G{\'o}rski}, {Hivon}, {Banday},
  {Wandelt}, {Hansen}, {Reinecke}, \& {Bartelmann}}]{Gorski2005}
{G{\'o}rski}, K.~M., {Hivon}, E., {Banday}, A.~J., {et~al.} 2005, \apj, 622,
  759

\bibitem[{{Gotz} {et~al.}(2022){Gotz}, {Mereghetti}, {Savchenko}, {Ferrigno},
  {Bozzo}, \& {IBAS Team}}]{Gotz2022}
{Gotz}, D., {Mereghetti}, S., {Savchenko}, V., {et~al.} 2022, GRB Coordinates
  Network, 32660, 1

\bibitem[{{Hastings}(1970)}]{Hastings1970}
{Hastings}, W.~K. 1970, Biometrika, 57, 97

\bibitem[{{Huang} {et~al.}(2022){Huang}, {Hu}, {Chen}, {Zha}, {Liu}, {Yao},
  {Cao}, \& {Experiment}}]{huang2022}
{Huang}, Y., {Hu}, S., {Chen}, S., {et~al.} 2022, GRB Coordinates Network,
  32677, 1

\bibitem[{{Knoll}(2000)}]{Knoll2000}
{Knoll}, G.~F. 2000, {Radiation detection and measurement} (Wiley)

\bibitem[{{Kozyrev} {et~al.}(2022){Kozyrev}, {Golovin}, {Litvak}, {Mitrofanov},
  {Sanin}, {Mgns/Bepicolombo Team}, {Hend/Mars Odyssey Team}, {Benkhoff},
  {Bepicolombo Team}, {Hurley}, {Ipn}, {Svinkin}, {Golenetskii}, {Frederiks},
  {Ridnaia}, {Lysenko}, {Cline}, {Konus-Wind Team}, {von Kienlin}, {Zhang},
  {Rau}, {Savchenko}, {Bozzo}, {Ferrigno}, {INTEGRAL SPI-ACS Grb Team},
  {Barthelmy}, {Cummings}, {Krimm}, {Palmer}, {Tohuvavohu}, {Swift-Bat Team},
  {Boynton}, {Fellows}, {Harshman}, {Enos}, {Starr}, {Gardner}, \& {Grs-Odyssey
  Grb Team}}]{Kozyrev2022}
{Kozyrev}, A.~S., {Golovin}, D.~V., {Litvak}, M.~L., {et~al.} 2022, GRB
  Coordinates Network, 32805, 1

\bibitem[{{Lesage} {et~al.}(2022){Lesage}, {Veres}, {Roberts}, {Burns},
  {Bissaldi}, \& {Fermi GBM Team}}]{lesage2022}
{Lesage}, S., {Veres}, P., {Roberts}, O.~J., {et~al.} 2022, GRB Coordinates
  Network, 32642, 1

\bibitem[{Lesage {et~al.}(2023)}]{lesage2023}
Lesage, S. {et~al.} 2023, in preparation

\bibitem[{{Malesani} {et~al.}(2023){Malesani}, {Levan}, {Izzo}, {de Ugarte
  Postigo}, {Ghirlanda}, {Heintz}, {Kann}, {Lamb}, {Palmerio}, {Salafia},
  {Salvaterra}, {Tanvir}, {Ag{\"u}{\'\i} Fern{\'a}ndez}, {Campana}, {Chrimes},
  {D'Avanzo}, {D'Elia}, {Della Valle}, {De Pasquale}, {Fynbo}, {Gaspari},
  {Gompertz}, {Hartmann}, {Hjorth}, {Jakobsson}, {Palazzi}, {Pian}, {Pugliese},
  {Ravasio}, {Rossi}, {Saccardi}, {Schady}, {Schneider}, {Sollerman},
  {Starling}, {Th{\"o}ne}, {van der Horst}, {Vergani}, {Watson}, {Wiersema},
  {Xu}, \& {Zafar}}]{Malesani2023}
{Malesani}, D.~B., {Levan}, A.~J., {Izzo}, L., {et~al.} 2023, arXiv e-prints,
  arXiv:2302.07891

\bibitem[{{Mazets} {et~al.}(2005){Mazets}, {Cline}, {Aptekar}, {Frederiks},
  {Golenetskii}, {Il'inskii}, \& {Pal'shin}}]{2005astro.ph..2541M}
{Mazets}, E.~P., {Cline}, T.~L., {Aptekar}, R.~L., {et~al.} 2005, arXiv
  e-prints [\eprint[arXiv]{astro-ph/0502541}]

\bibitem[{{M{\'e}sz{\'a}ros} {et~al.}(2022){M{\'e}sz{\'a}ros}, {P{\'a}l},
  {Werner}, {R{\'\i}pa}, {Ohno}, {Cs{\'a}k}, {Kapus}, {Frajt}, {Hudec},
  {Rezenov}, \& {Han{\'a}k}}]{meszaros2022}
{M{\'e}sz{\'a}ros}, L., {P{\'a}l}, A., {Werner}, N., {et~al.} 2022, in Society
  of Photo-Optical Instrumentation Engineers (SPIE) Conference Series, Vol.
  12181, Society of Photo-Optical Instrumentation Engineers (SPIE) Conference
  Series, ed. J.-W.~A. {den Herder}, S.~{Nikzad}, \& K.~{Nakazawa}, 121811L

\bibitem[{{Mitchell} {et~al.}(2022){Mitchell}, {Phlips}, \&
  {Johnson}}]{Mitchell2022}
{Mitchell}, L.~J., {Phlips}, B.~F., \& {Johnson}, W.~N. 2022, GRB Coordinates
  Network, 32746, 1

\bibitem[{{Oates}(2023)}]{Oates2023}
{Oates}, S. 2023, Universe, 9, 113

\bibitem[{{O'Connor} {et~al.}(2023){O'Connor}, {Troja}, {Ryan}, {Beniamini},
  {van Eerten}, {Granot}, {Dichiara}, {Ricci}, {Lipunov}, {Gillanders}, {Gill},
  {Moss}, {Anand}, {Andreoni}, {Becerra}, {Buckley}, {Butler}, {Cenko},
  {Chasovnikov}, {Durbak}, {Francile}, {Hammerstein}, {van der Horst},
  {Kasliwal}, {Kouveliotou}, {Kutyrev}, {Lee}, {Srinivasaragavan}, {Topolev},
  {Watson}, {Yang}, \& {Zhirkov}}]{OConnor2023}
{O'Connor}, B., {Troja}, E., {Ryan}, G., {et~al.} 2023, arXiv e-prints,
  arXiv:2302.07906

\bibitem[{{Paciesas} {et~al.}(1999){Paciesas}, {Meegan}, {Pendleton}, {Briggs},
  {Kouveliotou}, {Koshut}, {Lestrade}, {McCollough}, {Brainerd}, {Hakkila},
  {Henze}, {Preece}, {Connaughton}, {Kippen}, {Mallozzi}, {Fishman},
  {Richardson}, \& {Sahi}}]{1999ApJS..122..465P}
{Paciesas}, W.~S., {Meegan}, C.~A., {Pendleton}, G.~N., {et~al.} 1999, \apjs,
  122, 465

\bibitem[{{P{\'a}l} {et~al.}(2020){P{\'a}l}, {Ohno}, {M{\'e}sz{\'a}ros},
  {Werner}, {Ripa}, {Frajt}, {Hirade}, {Hudec}, {Kapu{\v{s}}}, {Koleda},
  {Laszlo}, {Lipovsk{\'y}}, {Matake}, {{\r{A}} melko}, {Uchida}, {Cs{\'a}k},
  {Enoto}, {Frei}, {Fukazawa}, {Galg{\'o}czi}, {Hirose}, {Hisadomi},
  {Ichinohe}, {Kiss}, {Mizuno}, {Nakazawa}, {Odaka}, {Takahashi}, \&
  {Torigoe}}]{pal2020}
{P{\'a}l}, A., {Ohno}, M., {M{\'e}sz{\'a}ros}, L., {et~al.} 2020, in Society of
  Photo-Optical Instrumentation Engineers (SPIE) Conference Series, Vol. 11444,
  Space Telescopes and Instrumentation 2020: Ultraviolet to Gamma Ray, ed.
  J.-W.~A. {den Herder}, S.~{Nikzad}, \& K.~{Nakazawa}, 114444V

\bibitem[{{Piano} {et~al.}(2022){Piano}, {Verrecchia}, {Bulgarelli}, {Ursi},
  {Panebianco}, {Pittori}, {Longo}, {Parmiggiani}, {Tavani}, {Argan},
  {Cardillo}, {Casentini}, {Evangelista}, {Foffano}, {Menegoni}, {Lucarelli},
  {Addis}, {Baroncelli}, {di Piano}, {Fioretti}, {Fuschino}, {Romani},
  {Marisaldi}, {Pilia}, {Trois}, {Donnarumma}, {Giuliani}, {Tempesta}, \&
  {Agile Team}}]{Piano2022}
{Piano}, G., {Verrecchia}, F., {Bulgarelli}, A., {et~al.} 2022, GRB Coordinates
  Network, 32657, 1

\bibitem[{{Pillera} {et~al.}(2022){Pillera}, {Bissaldi}, {Omodei}, {La Mura},
  \& {Longo}}]{pillera2022}
{Pillera}, R., {Bissaldi}, E., {Omodei}, N., {La Mura}, G., \& {Longo}, F.
  2022, The Astronomer's Telegram, 15656, 1

\bibitem[{{Planck Collaboration} {et~al.}(2020){Planck Collaboration},
  {Aghanim}, {Akrami}, {Ashdown}, {Aumont}, {Baccigalupi}, {Ballardini},
  {Banday}, {Barreiro}, {Bartolo}, {Basak}, {Battye}, {Benabed}, {Bernard},
  {Bersanelli}, {Bielewicz}, {Bock}, {Bond}, {Borrill}, {Bouchet}, {Boulanger},
  {Bucher}, {Burigana}, {Butler}, {Calabrese}, {Cardoso}, {Carron},
  {Challinor}, {Chiang}, {Chluba}, {Colombo}, {Combet}, {Contreras}, {Crill},
  {Cuttaia}, {de Bernardis}, {de Zotti}, {Delabrouille}, {Delouis}, {Di
  Valentino}, {Diego}, {Dor{\'e}}, {Douspis}, {Ducout}, {Dupac}, {Dusini},
  {Efstathiou}, {Elsner}, {En{\ss}lin}, {Eriksen}, {Fantaye}, {Farhang},
  {Fergusson}, {Fernandez-Cobos}, {Finelli}, {Forastieri}, {Frailis},
  {Fraisse}, {Franceschi}, {Frolov}, {Galeotta}, {Galli}, {Ganga},
  {G{\'e}nova-Santos}, {Gerbino}, {Ghosh}, {Gonz{\'a}lez-Nuevo}, {G{\'o}rski},
  {Gratton}, {Gruppuso}, {Gudmundsson}, {Hamann}, {Handley}, {Hansen},
  {Herranz}, {Hildebrandt}, {Hivon}, {Huang}, {Jaffe}, {Jones}, {Karakci},
  {Keih{\"a}nen}, {Keskitalo}, {Kiiveri}, {Kim}, {Kisner}, {Knox},
  {Krachmalnicoff}, {Kunz}, {Kurki-Suonio}, {Lagache}, {Lamarre}, {Lasenby},
  {Lattanzi}, {Lawrence}, {Le Jeune}, {Lemos}, {Lesgourgues}, {Levrier},
  {Lewis}, {Liguori}, {Lilje}, {Lilley}, {Lindholm}, {L{\'o}pez-Caniego},
  {Lubin}, {Ma}, {Mac{\'\i}as-P{\'e}rez}, {Maggio}, {Maino}, {Mandolesi},
  {Mangilli}, {Marcos-Caballero}, {Maris}, {Martin}, {Martinelli},
  {Mart{\'\i}nez-Gonz{\'a}lez}, {Matarrese}, {Mauri}, {McEwen}, {Meinhold},
  {Melchiorri}, {Mennella}, {Migliaccio}, {Millea}, {Mitra},
  {Miville-Desch{\^e}nes}, {Molinari}, {Montier}, {Morgante}, {Moss}, {Natoli},
  {N{\o}rgaard-Nielsen}, {Pagano}, {Paoletti}, {Partridge}, {Patanchon},
  {Peiris}, {Perrotta}, {Pettorino}, {Piacentini}, {Polastri}, {Polenta},
  {Puget}, {Rachen}, {Reinecke}, {Remazeilles}, {Renzi}, {Rocha}, {Rosset},
  {Roudier}, {Rubi{\~n}o-Mart{\'\i}n}, {Ruiz-Granados}, {Salvati}, {Sandri},
  {Savelainen}, {Scott}, {Shellard}, {Sirignano}, {Sirri}, {Spencer},
  {Sunyaev}, {Suur-Uski}, {Tauber}, {Tavagnacco}, {Tenti}, {Toffolatti},
  {Tomasi}, {Trombetti}, {Valenziano}, {Valiviita}, {Van Tent}, {Vibert},
  {Vielva}, {Villa}, {Vittorio}, {Wandelt}, {Wehus}, {White}, {White},
  {Zacchei}, \& {Zonca}}]{Planck_Collaboration2020}
{Planck Collaboration}, {Aghanim}, N., {Akrami}, Y., {et~al.} 2020, \aap, 641,
  A6

\bibitem[{{Terasawa} {et~al.}(2005){Terasawa}, {Tanaka}, {Takei}, {Kawai},
  {Yoshida}, {Nomoto}, {Yoshikawa}, {Saito}, {Kasaba}, {Takashima}, {Mukai},
  {Noda}, {Murakami}, {Watanabe}, {Muraki}, {Yokoyama}, \&
  {Hoshino}}]{2005Natur.434.1110T}
{Terasawa}, T., {Tanaka}, Y.~T., {Takei}, Y., {et~al.} 2005, \nat, 434, 1110

\bibitem[{{Tiengo} {et~al.}(2023){Tiengo}, {Pintore}, {Vaia}, {Filippi},
  {Sacchi}, {Esposito}, {Rigoselli}, {Mereghetti}, {Salvaterra}, {Siljeg},
  {Bracco}, {Bosnjak}, {Jelic}, \& {Campana}}]{Tiengo2023}
{Tiengo}, A., {Pintore}, F., {Vaia}, B., {et~al.} 2023, arXiv e-prints,
  arXiv:2302.11518

\bibitem[{{Ursi} {et~al.}(2022){Ursi}, {Panebianco}, {Pittori}, {Verrecchia},
  {Longo}, {Parmiggiani}, {Tavani}, {Argan}, {Cardillo}, {Casentini},
  {Evangelista}, {Foffano}, {Menegoni}, {Piano}, {Lucarelli}, {Addis},
  {Baroncelli}, {Bulgarelli}, {di Piano}, {Fioretti}, {Fuschino}, {Romani},
  {Marisaldi}, {Pilia}, {Trois}, {Donnarumma}, {Giuliani}, {Tempesta}, \&
  {Agile Team}}]{Ursi2022}
{Ursi}, A., {Panebianco}, G., {Pittori}, C., {et~al.} 2022, GRB Coordinates
  Network, 32650, 1

\bibitem[{{Veres} {et~al.}(2022){Veres}, {Burns}, {Bissaldi}, {Lesage},
  {Roberts}, \& {Fermi GBM Team}}]{veres2022}
{Veres}, P., {Burns}, E., {Bissaldi}, E., {et~al.} 2022, GRB Coordinates
  Network, 32636, 1

\bibitem[{{von Kienlin} {et~al.}(2020){von Kienlin}, {Meegan}, {Paciesas},
  {Bhat}, {Bissaldi}, {Briggs}, {Burns}, {Cleveland}, {Gibby}, {Giles},
  {Goldstein}, {Hamburg}, {Hui}, {Kocevski}, {Mailyan}, {Malacaria},
  {Poolakkil}, {Preece}, {Roberts}, {Veres}, \&
  {Wilson-Hodge}}]{2020ApJ...893...46V}
{von Kienlin}, A., {Meegan}, C.~A., {Paciesas}, W.~S., {et~al.} 2020, \apj,
  893, 46

\bibitem[{{{\v{R}}{\'\i}pa} {et~al.}(2022{\natexlab{a}}){{\v{R}}{\'\i}pa},
  {P{\'a}l}, {Ohno}, {Werner}, {M{\'e}sz{\'a}ros}, {Cs{\'a}k},
  {Daf{\v{c}}{\'\i}kov{\'a}}, {D{\'a}niel}, {Dud{\'a}{\v{s}}}, {Frajt},
  {Han{\'a}k}, {Hudec}, {Junas}, {Kapu{\v{s}}}, {Kasal}, {Koleda}, {Laszlo},
  {Lipovsky}, {M{\"u}nz}, {Rezenov}, {{\r{A}} melko}, {Svoboda}, {Takahashi},
  {Topinka}, {Urbanec}, {Breuer}, {Enoto}, {Frei}, {Fukazawa}, {Galg{\'o}czi},
  {Hroch}, {Ichinohe}, {Kiss}, {Matake}, {Mizuno}, {Nakazawa}, {Odaka}, {Poon},
  {Uchida}, \& {Uchida}}]{ripa2022a}
{{\v{R}}{\'\i}pa}, J., {P{\'a}l}, A., {Ohno}, M., {et~al.} 2022{\natexlab{a}},
  in Society of Photo-Optical Instrumentation Engineers (SPIE) Conference
  Series, Vol. 12181, Space Telescopes and Instrumentation 2022: Ultraviolet to
  Gamma Ray, ed. J.-W.~A. {den Herder}, S.~{Nikzad}, \& K.~{Nakazawa}, 121811K

\bibitem[{{{\v{R}}{\'\i}pa} {et~al.}(2022{\natexlab{b}}){{\v{R}}{\'\i}pa},
  {Pal}, {Werner}, {Ohno}, {Takahashi}, {Meszaros}, {Csak}, {Dafcikova},
  {Munz}, {Husarikova}, {Breuer}, {Topinka}, {Hroch}, {Urbanec}, {Kasal},
  {Povalac}, {Hudec}, {Kapus}, {Frajt}, {Laszlo}, {Koleda}, {Smelko}, {Hanak},
  {Lipovsky}, {Galgoczi}, {Uchida}, {Poon}, {Matake}, {Uchida}, {Bozoki},
  {Dalya}, {Enoto}, {Frei}, {Friss}, {Fukazawa}, {Hirose}, {Hisadomi},
  {Ichinohe}, {Kapas}, {Kiss}, {Mizuno}, {Nakazawa}, {Odaka}, {Takatsy},
  {Torigoe}, {Kogiso}, {Yoneyama}, {Moritaki}, {Kano}, \& {GRBAlpha
  Collaboration.}}]{ripa2022b}
{{\v{R}}{\'\i}pa}, J., {Pal}, A., {Werner}, N., {et~al.} 2022{\natexlab{b}},
  GRB Coordinates Network, 32685, 1

\bibitem[{{Werner} {et~al.}(2018){Werner}, {{\v{R}}{\'\i}pa}, {P{\'a}l},
  {Ohno}, {Tarcai}, {Torigoe}, {Tanaka}, {Uchida}, {M{\'e}sz{\'a}ros},
  {Galg{\'o}czi}, {Fukazawa}, {Mizuno}, {Takahashi}, {Nakazawa},
  {V{\'a}rhegyi}, {Enoto}, {Odaka}, {Ichinohe}, {Frei}, \& {Kiss}}]{werner2018}
{Werner}, N., {{\v{R}}{\'\i}pa}, J., {P{\'a}l}, A., {et~al.} 2018, in Society
  of Photo-Optical Instrumentation Engineers (SPIE) Conference Series, Vol.
  10699, Space Telescopes and Instrumentation 2018: Ultraviolet to Gamma Ray,
  ed. J.-W.~A. {den Herder}, S.~{Nikzad}, \& K.~{Nakazawa}, 106992P

\bibitem[{{Williams} {et~al.}(2023){Williams}, {Kennea}, {Dichiara},
  {Kobayashi}, {Iwakiri}, {Beardmore}, {Evans}, {Heinz}, {Lien}, {Oates},
  {Negoro}, {Cenko}, {Buisson}, {Hartmann}, {Jaisawal}, {Kuin}, {Lesage},
  {Page}, {Parsotan}, {Pasham}, {Sbarufatti}, {Siegel}, {Sugita}, {Younes},
  {Ambrosi}, {Arzoumanian}, {Bernardini}, {Campana}, {Capalbi}, {Caputo},
  {D'Ai}, {D'Avanzo}, {D'Elia}, {De Pasquale}, {Eyles-Ferris}, {Ferrara},
  {Gendreau}, {Gropp}, {Kawai}, {Klingler}, {Laha}, {Melandri}, {Mihara},
  {Moss}, {O'Brien}, {Osborne}, {Palmer}, {Perri}, {Serino}, {Sonbas},
  {Stamatikos}, {Starling}, {Tagliaferri}, {Tohuvavohu}, {Zane}, \&
  {Ziaeepour}}]{Williams2023}
{Williams}, M.~A., {Kennea}, J.~A., {Dichiara}, S., {et~al.} 2023, arXiv
  e-prints, arXiv:2302.03642

\bibitem[{{Xiao} {et~al.}(2022){Xiao}, {Krucker}, \& {Daniel}}]{Xiao2022}
{Xiao}, H., {Krucker}, S., \& {Daniel}, R. 2022, GRB Coordinates Network,
  32661, 1

\bibitem[{{Yonetoku} {et~al.}(2010){Yonetoku}, {Murakami}, {Tsutsui},
  {Nakamura}, {Morihara}, \& {Takahashi}}]{Yonetoku2010}
{Yonetoku}, D., {Murakami}, T., {Tsutsui}, R., {et~al.} 2010, \pasj, 62, 1495

\end{thebibliography}

\end{document}